# Software Fairness: An Analysis and Survey

EZEKIEL SOREMEKUN, MIKE PAPADAKIS, MAXIME CORDY, and YVES LE TRAON, SnT, University of Luxembourg, Luxembourg

**ABSTRACT** In the last decade, researchers have studied fairness as a software property. In particular, how to engineer fair software systems? This includes specifying, designing, and validating fairness properties. However, the landscape of works addressing bias as a software engineering concern is unclear, i.e., techniques and studies that analyze the fairness properties of learning-based software. In this work, we provide a clear view of the state-of-the-art in software fairness analysis. To this end, we collect, categorize and conduct in-depth analysis of 164 publications investigating the fairness of learning-based software systems. Specifically, we study the evaluated fairness measure, the studied tasks, the type of fairness analysis, the main idea of the proposed approaches and the access level (e.g., black, white or grey box). Our findings include the following: (1) Fairness concerns (such as fairness specification and requirements engineering) are under-studied; (2) Fairness measures such as conditional, sequential and intersectional fairness are under-explored; (3) Unstructured datasets (e.g., audio, image and text) are barely studied for fairness analysis; and (4) Software fairness analysis techniques hardly employ white-box, in-processing machine learning (ML) analysis methods. In summary, we observed several open challenges including the need to study intersectional/sequential bias, policy-based bias handling and human-in-the-loop, socio-technical bias mitigation.

## 1 INTRODUCTION

*Software fairness* is a property of learning-based systems which aims to ensure that the software does not exhibit biases [131]. Given a set of inputs, a fair software should not result in discriminatory outputs or behaviors for inputs relating to certain groups or individuals. In essence, the goal is to ensure that software systems exhibit fair behavior for all inputs that are similar for the task-at-hand. For instance, *discriminatory inputs*, inputs that are similar for a task but only differ in *sensitive attributes* (e.g., gender, race or age), should produce similar outputs or induce similar program behaviors [56].

Specifically, the goal of *software fairness analysis* is to ensure a given system produces the same results or exhibit similar behaviors for a number of *discriminatory inputs*. This is important to detect, expose, diagnose and mitigate bias (i.e., discrimination) in software systems. As an example, consider a sentiment analyzer software that determines the emotional state or situation in a text, which outputs either a positive emotion (e.g., text describing excitement) or negative emotion (e.g., text portraying sadness or anger). Figure 1 shows an example of a bias in such a system, where

Authors' address: Ezekiel Soremekun, ezekiel.soremekun@uni.lu; Mike Papadakis, michail.papadakis@uni.lu; Maxime Cordy, maxime.cordy@uni.lu; Yves Le Traon, Yves.LeTraon@uni.lu, SnT, University of Luxembourg, 6, rue Richard Coudenhove-Kalergi, Luxembourg, Luxembourg, Luxembourg, L-1359.

the output (sentiment) is different when the gender of the noun in the text is a "man" compared to a "woman". This is an illustrative example of (gender) bias found in real-world natural language processing (NLP) systems [11, 95, 121].

Several researchers have studied software fairness analysis, with the aim to address a fundamental question: How to engineer fair software systems? Such works consider fairness as a non-functional software property and a software engineering (SE) concern. This includes work studying how to specify [14, 117], test [56], and mitigate [34] fairness properties in software systems. However, despite several works on software fairness analysis, it is difficult to understand the state of research practice: Specifically, what fairness concerns have been addressed? How have they been addressed? What are the open problems and challenges?

In this paper, we aim to provide a clear view of the state-of-the-art in software fairness analysis, i.e., techniques and studies that analyse the fairness properties of learning-based software. This paper aims to analyze the trends in software fairness analysis, the available techniques, the focus of the research community, the problems that have been addressed and the open research problems. To this end, we perform a systematic analysis of the literature, where we studied 164 papers studying *fairness as a software property*. These papers are mostly published in fairness-related or software engineering (SE) venues, as well as venues focused on machine learning (ML), artificial intelligence (AI), security, computer vision (CV) and natural language processing (NLP). Particularly, we conduct an in-depth study of the set of publications that explore fairness with the lens of software engineering, e.g., in terms of software quality control, requirement engineering, design and development. We then characterize several factors encapsulated in these research papers, including the evaluated fairness measure (e.g., individual, group, causal or conditional fairness), the studied tasks (e.g., credit rating, CV, NLP), the type of fairness analysis (e.g., testing or mitigation), the main idea of the proposed approach and the level of access (e.g., black, white or grey box).

Overall, we observe that the research community is facing several open challenges in addressing the following fairness concerns: compounding or intersectional bias, verification of fairness properties, sequential bias, equity-based bias handling and socio-technical solutions to bias. The key findings of this work includes the following:

- Software Fairness analysis is mostly performed to validate or mitigate biases, i.e., the focus of the community has been to test and reduce unfair program behaviors, other fairness concerns (such as requirements engineering and verification) are under-studied;
- The most studied fairness measure include individual, group and causal fairness, measures such as conditional, sequential and intersectional fairness remain under-explored;
- The most examined tasks involve structured datasets (such as the adult income dataset), while unstructured datasets (e.g., audio, image and text) are barely studied in the SE community;
- The most employed techniques in the SE community are mutation (e.g., input perturbation) and specification (e.g., using input templates, schemas or grammars) -based techniques, other approaches such as in-processing machine learning (ML) analysis methods are hardly employed for software fairness analysis;
- Most approaches support the analysis of an atomic attribute (e.g. race, or gender), very few approaches study the combination or sequence of attributes (e.g., race × gender), the compounding effect of multiple instances of an attribute, or complex attributes (such as non-binary gender);
- We found little or very few works tackling the fairness concerns like fairness test metrics/adequacy, automatic repair of biased classifiers or time-based fairness concerns (e.g., sequential or regression fairness bugs).

The rest of this paper is structured as follows: We provide background on software fairness in section 2. We discuss the process of collecting/analyzing publications in section 3, and section 4 highlights our research questions. In section 5, we provide in-depth analysis of our findings, and section 6 discusses the limitations and threats to validity of our study. Finally, we conclude with discussions of open research challenges in software fairness analysis (section 7).

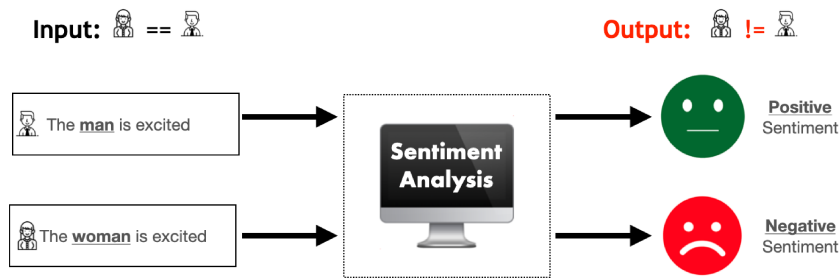

Fig. 1. Example of Individual Fairness Violation using a Sentiment Analysis AI System

## 2 BACKGROUND

We provide background on software fairness, with the definition of terms employed throughout the paper, illustrative examples and a discussion of closely related works.

### 2.1 Definition of Terms

We describe the main terms used in the rest of this paper, and the context in which they apply to our analysis and survey.

**Bias:** In this paper, we refer to *bias* in terms of algorithmic bias, specifically, bias occurs when a learning-based software system *systematically* and *unfairly* discriminate against certain individuals or groups of individuals in favour of others [54]. Algorithmic bias causes discrimination against certain people and can lead to real-world harms in terms of the representation or allocation of resources to such individuals or groups [37].

**Software Fairness:** There has been a significant work in understanding and defining *fairness* as a software behavior [131]. In this work, we examine *fairness as a software property*. The aim is to focus on works examining *software fairness via the lens of software engineering*, e.g., how to engineer bias-aware software or prevent bias in software systems. To this end, we study papers that study fairness as a SE concern including in terms of a requirements engineering [17], software quality control [56], software design [74], and software verification [7].

**Fairness Metrics:** Verma and Rubin [131] provides comprehensive definitions and categories of several fairness metrics employed in the literature. Figure 1 and Figure 2 exemplify the two most popular metrics, namely *individual fairness* and *group fairness*. In the following, we define these two metrics.

*Individual fairness:* Dwork et al. [44] provides a comprehensive definition of individual fairness. Individual fairness means that the software should treat similar individuals similarly. Consider the sentiment analysis system in Figure 1, it violates individual fairness since it treats a text with a "man" as a *noun* differently from that with a "woman". This metric requires that the individuals should be similar for the purposes of the respective task and the outcomes should have similar distributions. Formally, individual fairness is a violation of the following condition:

$$\left| f(a) - f(a') \right| \leq \tau \tag{1}$$

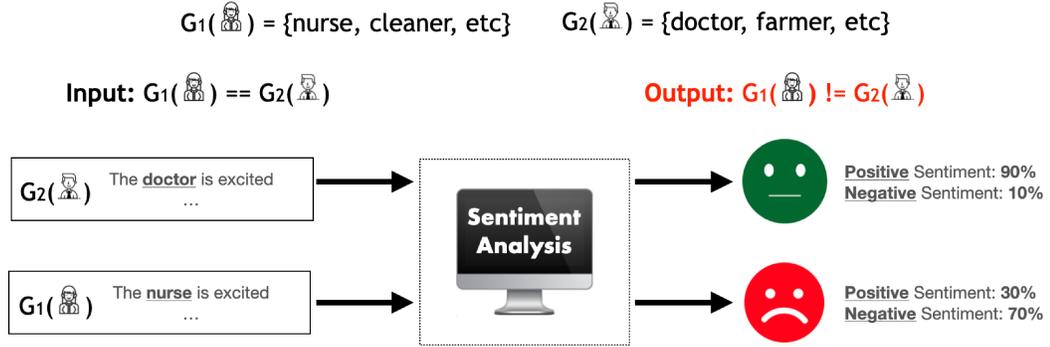

Fig. 2. Example of Group Fairness Violation using a Sentiment Analysis AI System

Here, $a$ and $a'$ are similar individuals (inputs), $f$ is a software esystem (e.g., automated classifier) and $\tau$ is some threshold which is chosen using the inputs and the model as context. In our example (Figure 1), the text containing "man" or "woman" should be treated similarly as individuals, however the output shows that the system provides different outcomes for the similar inputs.

*Group fairness:* A system satisfies group fairness if subjects in the protected and unprotected groups have equal probability of being assigned a particular outcome [131]. This fairness metric aims to ensure that two or more groups are treated similarly. Consider the sentiment analysis system in Figure 2, it violates group fairness since it treats texts containing nouns belonging to *male-biased occupations* (e.g., doctor, farmers etc) differently from similar texts that contain *female-biased occupations* (e.g., nurse, cleaner, etc). Formally, group fairness is maintained if the following condition is true:

$$Pr(f(a) = +|A = a) = Pr(f(b) = +|A = b) \; \forall a, b \in A \tag{2}$$

Given equivalent inputs from different groups $a$ and $b$, the aforementioned definition checks for the equivalence of the outputs from software $f$. Here, the choice of a group is determined by random variable $A$ and the positive prediction rate is denoted by $+$. In this example, the groups are *male-biased female-biased occupations* and *female-biased female-biased occupations*, and the model produces significantly different distribution of positive and negative predictions for each group. As an example, a group fairness violation is that texts containing male-biased jobs are more (90%) likely to return a positive sentiment than a female-biased job (30%).

## 2.2 Related Work

In the last few years, several researchers have surveyed the problem of bias or fairness in learning-based software, but these surveys have been mostly *specialised*, i.e., they studied this problem for a specific domain, bias, or metric. Concretely, previous surveys on fairness analysis target a particular sub-domain (e.g., NLP [23], CV [46], ranking [139] or finance [106]), a specific sensitive attribute or bias (e.g., race [51]), or a specific fairness metric (e.g., causal fairness [97]). Thus, these surveys do not account for the advances in other domains, attributes and metrics. Other fairness surveys are either more general beyond fairness concerns or are focused on specific goals, e.g., general testing of ML systems for several properties [142], or providing a taxonomy for (sources of) bias in ML systems [98], respectively. More importantly, none of these papers have performed a systematic analysis of bias in the context of *software engineeering* or *the analysis of software fairness*. Specifically, studies and methods that *(automatically) evaluate software fairness* *properties in learning-based systems*. Besides, this paper (provides a comprehensive view of the literature w.r.t. to SE, involving significantly more (SE-related) publications than the aforementioned specialized surveys. In the following we discuss the closely related work to this paper, in particular the published surveys in this area.

There are a few *domain-specific* surveys of fairness where researchers have surveyed fairness issues in a specific sub-domain of learning-based software, e.g., recommendation systems [45], ranking-based systems [139], sequential decision systems [146], NLP [23], CV [46], or financial services [106]. Blodgett et al. [23] comprehensively analyzed 146 papers addressing fairness of NLP systems, especially quantitative techniques for measuring or mitigating bias. The authors found that almost all surveyed papers are poorly matched to their motivations and do not engage with the relevant literature outside of NLP. The authors recommend that fairness analysis in NLP systems should be based on characterizing system behaviors that are harmful, by centering bias mitigation around people and their experiences. Fabbrizzi et al. [46] provides a survey on bias in visual datasets, i.e., for CV applications. The authors describe different biases in visual datasets, the proposed methods for detecting and measuring these biases and existing bias-aware datasets. The authors concluded that there is no bias-free dataset and detecting biases in visual datasets is an open problem and recommend a checklist to help practitioners spot different biases in their dataset, in order to make bias explicit. Meanwhile, Zehlike et al. [139] studied the application of fairness-enhancing interventions in ranking algorithms, especially examining the literature on incorporating fairness requirements into algorithmic rankers. The authors surveyed papers from several venues, including data management, algorithms, information retrieval, and recommendation systems. The goal of the survey is to provide a new framework that unifies fairness mitigation objectives and ranking requirements, such that it allows to examine the trade-off between both goals. Ekstrand et al. [45] studied how algorithmic fairness issues applies to information retrieval and recommendation systems, especially addressing how to translate algorithmic fairness from classification, scoring, and ranking settings into recommendation and information retrieval settings.[1] In addition, Zhang and Liu [146] provides a literature review of fairness in sequential learning-based decision-making systems, i.e., systems where decision-making are not a one time event, but rather occur in a sequential nature, such that decisions made in the past may have an impact on future data. Unlike these papers, this work is not *domain-specific*, instead, we study the problem of fairness analysis across several (sub-)domains, including all of the aforementioned domains.

Moreover, other surveys on fairness are *bias-specific* or *metric-specific*, they either focus on a specific sensitive attribute (e.g., race) [51], or a specific fairness metric (e.g., causal fairness) [97], respectively. For instance, Field et al. [51] provides a survey of race-related bias in the stages of NLP model development. The authors surveyed 79 papers with the goal of understanding the gaps between *race-related bias* analysis in NLP and other related fields. The authors found that race has been siloed as a niche topic in the NLP community and often ignored in many NLP tasks. The authors also emphasize the need for racial inclusion in NLP research. In addition, Makhlouf et al. [97] study the application of causality to address the problem of fairness, by studying papers that examine *causal fairness* properties and their applicability in real-world scenarios. The authors employed *identifiability theory* to determine the criteria for the real-world applicability of *causal fairness*. However, in this work, we examine papers examining several sensitive attributes, biases and fairness metrics. In particular, unlike these metric or bias -specific surveys, we do not focus on papers examining a single type of bias or fairness metric. Instead, we evaluate the literature across different biases and fairness metrics, including race and causal fairness, respectively.

---

[1] https://fair-ia.ekstrandom.net/

A few researchers have conducted surveys of the literature on software fairness, albeit mostly targeting fair prediction in machine learning [31, 55, 98, 101], general ML testing [142] or fairness notions across domains [68] or in specific domains, e.g., concurrent systems [85]. For instance, Gajane and Pechenizkiy [55] studied the *formalization* of fairness in the machine learning literature for prediction tasks, especially examining how this relates to the notions of distributive justice in the social science literature. In their study, the authors proposed that two notions of distributive justice be formalised for fairness in ML, namely *equality of resources* and *equality of capability of functioning*. Likewise, Ntoutsi et al. [101] provides an introductory survey on the technical challenges and available solutions to bias in data-driven AI, with a focus on the legal grounds for bias challenges and the societal implications of these solutions. Mehrabi et al. [98] examined the fairness issues in real-world applications, specifically investigating the different sources of bias in AI systems and providing a taxonomy of fairness definitions in the ML community. Similarly, Caton and Haas [31] provide an overview of fairness mitigation approaches for ML, by categorising mitigation techniques into several stages in the ML model development pipeline. The authors highlight 11 mitigation methods categorised into three areas, namely pre-processing, in-processing, and post-processing methods. In addition, Zhang et al. [142] surveyed the testing of several properties in machine learning (ML), including fairness properties. In contrast to Zhang et al. [142], we investigate fairness improvement beyond automated testing approaches, for instance, we also examine papers examining its formalization, empirical evaluation, detection and improvement. Meanwhile Hutchinson and Mitchell [68] and [85] studied the notions of fairness properties across domains, and for concurrent systems, respectively. Notably, Hutchinson and Mitchell [68] surveyed the history of the definitions of fairness properties over the last 50 years across multiple disciplines, including education, hiring, and machine learning. The authors compared past and current notions of fairness along several dimensions, including the fairness criteria, the purpose of the criteria (e.g., testing) and how it relates to the mathematical method for measuring fairness (e.g., classification, regression) and people (individuals, groups, and subgroups). Unlike the aforementioned works, our survey of fairness is *more general*, we study software fairness beyond advances in specialised ML communities. In this work, we additionally examine several papers from security, CHI, PL, CV and NLP venues.

Overall, we provide a systematic literature review of the analysis of software fairness properties for learning-based systems. To the best of our knowledge, this work is the only systematic literature review concerned with the *analysis of fairness a (non-functional) software property of learning-based systems*. We are not aware of any other survey that focuses on this research area, i.e., software fairness analysis, especially providing a comprehensive survey of its formalization, testing, diagnosis and mitigation across several fairness metrics, biases and domains.

## 3 METHODOLOGY

The research methodology employed in this work is based on the methodology detailed in Kitchenham [83]. In the following, we provide the details of our research protocol:

(1) **Aim and Scope:** First, we define the goals of this work and the scope of works to be examined. Then, we define the scientific questions, the analysis protocol and the information relevant for our data analysis. To this end, we define the research questions (*see* section 4).
(2) **Detailed Information:** To analyse each paper in-depth, we identify the information necessary to answer all research questions, this includes information that allow us to categorise, understand and describe the problem addressed by each approach, and the technique or results provided by each paper. This informed the publication search (e.g., our focus venues,), as well as the keywords employed in the filtering process used in identifying relevant papers. Among several details, our interests include the studied fairness metrics and biases, the form of fairness analysis (e.g., fairness testing) and the level of access (e.g., white, grey or black -box).

(3) **Publication Search:** We curated fairness-related publication via three means, (1) we searched the top venues in SE (such as ICSE, TSE and FSE), Programming Languages (PL) (e.g., PLDI and POPL), Security (e.g., CCS and Euro S & P), Artificial Intelligence (AI) (e.g., AAAI), ML (e.g., ICML, NeurIPS), CV (e.g., CVPR) and NLP (e.g., ACL, EMNLP), (2) we collected papers from fairness focused conferences and workshops such as FairWare, FAT, FATML, and FaaCT; and (3) we conducted a keyword guided publication search of paper respositories (such as ACM Digital library[2], IEEE Xplore Digital Library[3] and Google Scholar[4]) using popular fairness-related terms such as "bias", "fairness", "discrimination", etc. In total we merged all collected papers which amounted to 420 papers from 64 different venues.

(4) **Filtering:** To identify relevant publications we filter out publications that are not relevant to our goals and analysis. Specifically, we filter out papers that (1) do not conduct fairness analysis as a part of the software engineering process, i.e., papers that are not relevant to the requirements, design and quality control of AI or ML -based software systems; (2) we exclude papers that are not written in English language, duplicate papers, as well as short papers or extended abstracts; (3) we also exclude papers that analyze fairness for other systems, i.e., fairness for non-learning-based software systems (e.g., fairness of network systems); (4) we also exclude papers that are not focused on software fairness properties, e.g., papers studying other properties such as robustness, consistency, security or accuracy; (5) finally, we read the abstract of each paper and excluded papers that are not research papers studying software fairness, for instance, we excluded surveys, literature reviews and invited lectures. In total, we filtered out 256 papers and were left with 164 papers for our analysis.

Given the large number of collected papers, we designed a research protocol to analyse each paper and extract certain information from the papers. For each paper, we extract both *the metadata* of the paper, as well as the *detailed research information*. In terms of metadata, we extracted the the author details, the paper title, the year and venue of publication and the domain of the publication (e.g., SE, security or PL). For detailed research-relevant information, we studied the following details about the techniques and evaluation of the paper: the employed datasets, the studied fairness metrics, the studied biases (i.e., sensitive attributes, e.g., race), the form of analysis (e.g., fairness testing), the access level, the specific problem addressed, the main idea of the paper, the proposed solution, the resulting findings, as well as the strengths and weaknesses of proposed analysis/solution.

## 4 RESEARCH QUESTIONS

In this work, we examine several publications (164) that *analyse software fairness in learning-based systems*. Table 1 provides details of some of the collected publications. We analyse publications from different domains and venues, including SE, PL, security, AI//ML, CV and NLP (*see* Figure 4). Firstly, we examine the research advances and trends, we investigate the volume and categories of publications from 2010 till date (i.e., early 2022) (RQ1). Secondly, we investigate the purpose of these publications including the main idea of the proposed techniques as well as how SE researchers study software fairness as a software engineering task (RQ2). For instance, we examine if the aim of the proposed method or analysis is to formalize, test, mitigate or diagnose fairness issues in learning-based software systems. Next, we analyse the fairness measure studied in the papers, such as individual, group, causal or intersectional fairness (RQ3).

---

[2] https://dl.acm.org/
[3] https://ieeexplore.ieee.org/Xplore/home.jsp
[4] https://scholar.google.com/

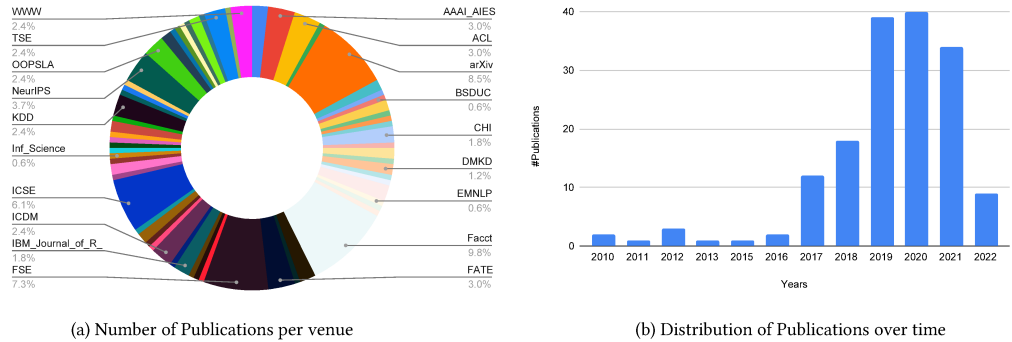

(a) Number of Publications per venue  (b) Distribution of Publications over time

Fig. 3. Distribution of Publication Venues and Year of Publications

We study the bias, i.e., the sensitive or protected attributes (e.g., gender or race) studied in the literature (RQ4). We also examine the tasks and datasets employed in the reviewed publications (RQ5). Finally, we investigate the tools available for software fairness analysis in the literature (RQ6).

Specifically, we aim to address the following research questions (RQs).

- **RQ1 Research Trends and Advances.** What is the research trend in software fairness analysis, in terms of the details of publications over the years, e.g., the volume, type, domain and venue of publications? What advances have been made over the years, especially in specific research areas (e.g., fairness validation and verification) and venues (e.g., SE and data-centric venues)?
- **RQ2 Purpose of Fairness Analysis.** What is the purpose of fairness analysis in the literature? What fairness problem is addressed or studied in the literature? What is the target or focus of the community in terms of fairness? What areas have been well-examined or un(der-)investigated in the research community?
- **RQ3 Fairness measure.** What are the fairness metrics analysed by the research community (e.g., individual, group or causal fairness)? What metrics are well-studied, under-studied or not investigated?
- **RQ4 Bias and Sensitive Attributes.** Which (societal) biases (e.g., age, race gender or religion) are investigated in the analysis of software fairness, especially w.r.t. to protected or sensitive attributes?
- **RQ5 Datasets and Tasks.** What datasets and tasks are employed for software fairness analysis? What tasks/datasets have been well or under-studied? What is the distribution of datasets per tasks?
- **RQ6 Tooling.** What are the available fairness analysis tool(kit)s, frameworks and libraries? What problems do these tools address, what are the analysis goals supported by available tools? Which analysis approaches are employed in the tools? What stage of model processing do these tools support, and what level of model access do they require?

## 5 RESEARCH FINDINGS

### 5.1 RQ1 Research Trends and Advances.

Let us analyze the volume of publications. We first examine the metadata of our publication corpus to determine the volume of publications in different domains and venues. For this analysis, we collected a corpus of 164 papers from 64 different venues and eight (8) different academic domains/fields. Table 1 and Figure 3 show the details of publication venues and the distribution of the publications in our corpus by venue. We are also interested in analysing the details of

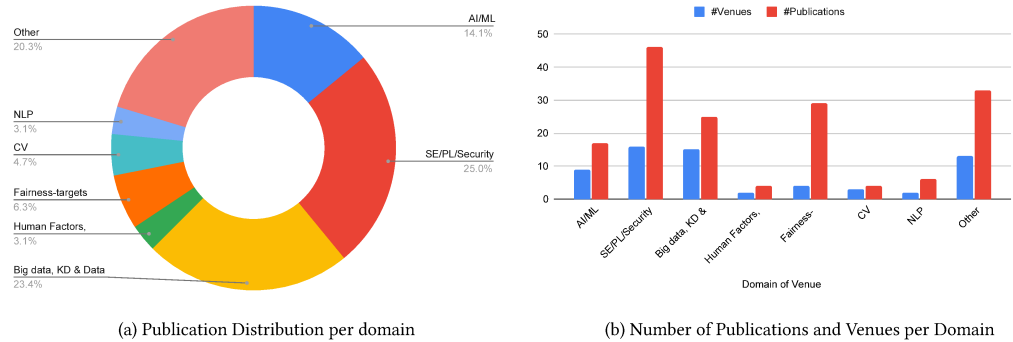

(a) Publication Distribution per domain    (b) Number of Publications and Venues per Domain

Fig. 4. Details of Publication domains

Table 1. Excerpt of Publication Details ("#" means "number of")

| Domain (#Pubs.) | Type | Venue | | Publications | | |
|---|---|---|---|---|---|---|
| | | #Venues | Sample Venue(s) | #Pubs | Example Publications | Years |
| **Software Engineering (SE), Programming Languages (PL) & Security** | *Conference* | 9 | ASE, CAV, EuroS&P, FSE, GECCO, ICSE, OOPSLA, TrustCom, ISSTA | 34 | [20, 35, 47, 56, 57, 65, 88, 126] | 2017-22 |
| | *Journal* | 4 | EMSE, JSS, RE, TSE | 7 | [11, 16, 17, 52, 120, 121, 145] | 2009-21 |
| | *Other* | 1 | ICSE-C | 1 | [6] | 2021 |
| **Natural language processing (NLP)** | *Conference* | 2 | ACL, EMNLP | 6 | [23, 24, 51, 107, 113, 147] | 2017-21 |
| **Artificial Intelligence (AI) & Machine Learning (ML)** | *Conference* | 8 | AAAI, AISTATS, ICML, NeurIPS, PMLR | 17 | [4, 82, 99, 140] [3, 30, 70, 78, 104, 135] | 2013-21 |
| | *Other* | 1 | HRLC | 1 | [146] | 2021 |
| **Computer Vision (CV)** | *Conference* | 2 | ICCV, CVPR | 3 | [80, 132, 133] | 2019-20 |
| | *Workshop* | 1 | ECCV | 1 | [138] | 2020 |
| **Fairness-targets** | *Conference* | 2 | AAAI-AIES, Facct | 21 | [9, 18, 22, 32, 53, 62, 79, 109, 118] | 2017-21 |
| | *Workshop* | 2 | FairWare, FATE | 8 | [12, 25, 40, 41, 67, 131, 134] | 2018-19 |
| **Big Data, Data Mining (DM), & Knowledge Discovery (KD)** | *Conference* | 7 | DMKD, ECML-PKDD, EDBT, KDD, ICDM, ICEDT, ICMD, LAK | 18 | [29, 75, 87, 149] [36, 49, 74, 77, 123] | 2010-21 |
| | *Journal* | 5 | Big Data, Inf. Science, JDIQ, KAIS, SIGMOD-Record | 5 | [1, 42, 73] [76, 112] | 2012-19 |
| | *Workshop* | 2 | BSDUC, KDD-XAI | 2 | [60, 108] | 2018-19 |
| **Human Factors & Usability** | *Conference* | 1 | CHI | 3 | [38, 63, 86] | 2019-21 |
| | *Journal* | 1 | IWC | 1 | [27] | 2016 |
| **Others** | *Conference* | 3 | CCCT, VAST, WWW | 6 | [28, 48, 61, 72, 84] | 2009-20 |
| | *Journal* | 6 | CACM, DGRP, Scientific-data, SSRN, IBM Journal of R & D | 10 | [15, 58, 69, 114, 115] | 2018-21 |
| | *Workshop* | 1 | CEUR Workshop | 1 | [122] | 2019 |
| | *Other* | 3 | arXiv, HRDAG, MS Tech. Report | 15 | [19, 39, 64, 71, 92, 111, 117] | 2019-21 |

the publications in terms of the type of venues (e.g., conference, journals and workshops) and the domain (or field) of publication venues (e.g., SE or AI/ML). Figure 4 and Figure 5 show the details of our publications in terms of domains and venue type, respectively.

***Volume and Domain of Publications:*** Our analysis showed that even though algorithmic fairness is studied across different research communities, *the study of fairness as a software property is mostly dominant in the SE and data-centric venues*. Notably, *the majority (about 48.3%) of all publications on software fairness are in the SE (e.g., FSE) and data-centric (i.e., Big data, knowledge discovery and data mining) venues (e.g., KDD).* Followed by typical top-tier AI/ML domains (e.g., AAAI and NeurIPS) which account for about 14.1% of all papers on software fairness. Figure 4(a) further illustrates that the top software engineering venues (e.g., ICSE, FSE, ISSTA, ASE, TSE, OOPSLA) account for about one in four (about 25.0% of) publications, with this commiunity accounting for most of the venues and publications (*see* Figure 4(b)). In addition, we observed that most papers in our corpus are published in three main venues, namely ICSE (6.1%), FSE

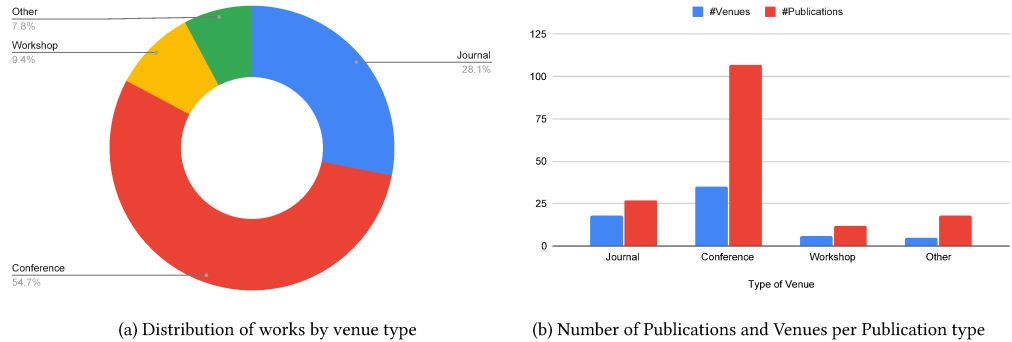

(a) Distribution of works by venue type    (b) Number of Publications and Venues per Publication type

Fig. 5. Details of the type of Publication Venue (e.g., Conference or Journal)

(7.9%) or FaccT (9.8%). Other popular venues include Big Data, Knowledge Discovery and Data Mining domains (23.4%), as well as AI/ML domains (14.1%) (*see* Figure 4). Likewise, fairness-focused venues (such as FaccT, FATE and AAAI AIES) account for about 6.3% of all publications. Meanwhile, more specialised venues (e.g., ECCV and CAV) had the least number of works on software fairness. This shows that even though research works on software fairness analysis are published across different domains, most publications are in SE venues. This demonstrates the growing interest in software fairness within the SE community. In particular, the need to apply SE processes, methods and tools to study the implications of fairness measures in practice has become paramount in the SE community.

> *Majority (48%) of software fairness analysis are published in typical SE venues and Data-centric venues, this is followed by top-tier AI/ML venues which account for about 14.1% of all publications.*

**Publication Trends over time:** Examining the volume of publications over the years, we observed that *the number of publications in software fairness analysis has been steadily increasing over the last decade.* Figure 3(b) highlights the trend over the last 13 years, it shows that the number of publications in software analysis has been steadily increasing over time, particularly in the last half decade.[5] This trend signifies the growing research interests in fairness analysis in the SE research community. Particularly, a major surge in publications can be observed starting from 2017 till now (i.e., early 2022). This is following the publication of Galhotra et al. [56] which *first formalizes causal fairness as a non-functional property of (learning-based) software systems.* Indeed, almost all software fairness papers analysied in this paper (152 papers, 92.68%) were published starting from 2017. This demonstrates the growing interests and increasing number of research output in this area over the years.

**Type of Publication Venues:** Figure 5 illustrates the distribution of the type of publication venues in our corpus. We observed that the majority (54.7%) of the publications on fairness analysis are in conference proceedings (*see* Figure 5 (a)). Figure 5(b) also shows that conferences are the most popular venues for software fairness publications, especially top-tier venues like ICSE and FSE, as well as popular conferences focused on Fairness (e.g., FaccT). Journal publications account for about 28.1% of all publications, with TSE, EMSE and JSS being the most popular journal venues. The most popular workshop venue for fairness analysis is FairWare. These findings show that software fairness has (more recently) become an important research area for top-tier SE conferences and journals.

---

[5]Note that the data point for the volume of publications in 2022 was obtained only early in the year, i.e., representing publications for only the first quarter of 2022.

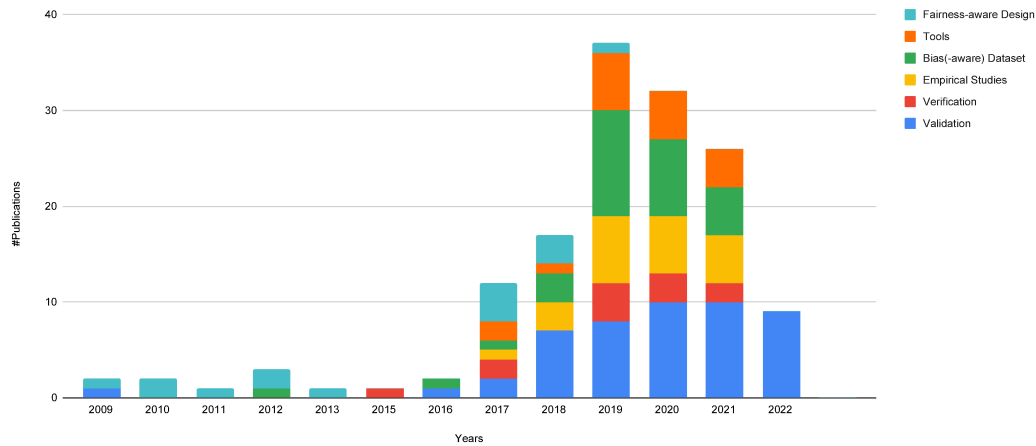

Fig. 6. Detailed Publication Trend by Year

> *Most (92.68%) papers analysing software fairness were published since 2017, and majority of these works (54.7%) were published at SE-focused conferences (e.g., ICSE and FSE).*

***Advances:*** We analyse the advances made in the analysis of software fairness by focusing on six major areas of fairness concern and their advances over time. Specifically, we analyse the trends in engineering concerns in software fairness such as validation, verification, design, empirical studies, tooling and datasets. Figure 6 provides a detailed overview of the trend and advances in publications for all of these six concerns. Figure 7 and Figure 8 also provide detailed distribution of publications in each area over the years.

Generally, the number of publications has increased over time for all six SE areas. However, there has been major increase in publications in the area of fairness validation (*see* Figure 7(a)), bias-aware datasets and empirical studies (*see* (*see* Figure 8 (b) and (c)). This is evident by the number of publications (about eight to 12 publications yearly) in these areas in the last half decade (since 2018). In addition, analysing the span of publications also show that fairness-related validation and datasets have been studied for about seven consecutive years, while empirical studies in fairness have only recently become prominent (i.e., in the last 5 years).

Meanwhile, other SE concerns such as the verification, tooling and design of fair software systems have received little attention. These areas have seen fewer (four to six maximum) publications in the last years. However, the span of publications in the design of fair software is larger, with at least a 10 year spread (*see* Figure 7(c)), showing that there has been a steady interest in this area over the years. This implies the need to investigate and examine approaches and methods to address the under-studied areas, especially the verification and tooling of fairness-aware software systems.

> *There has been advances in several areas of software fairness over the years, but most consecutive publications has been in fair learning and validation, with up to 10 yearly publications in the last five years.*

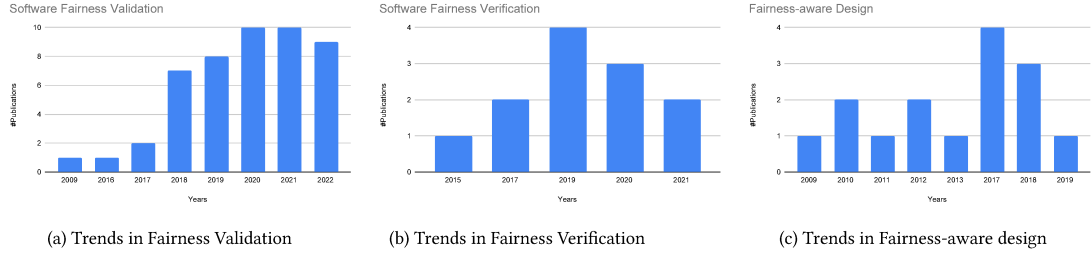

(a) Trends in Fairness Validation  (b) Trends in Fairness Verification  (c) Trends in Fairness-aware design

Fig. 7. Details of trends in Fairness Verification, Validation and Design

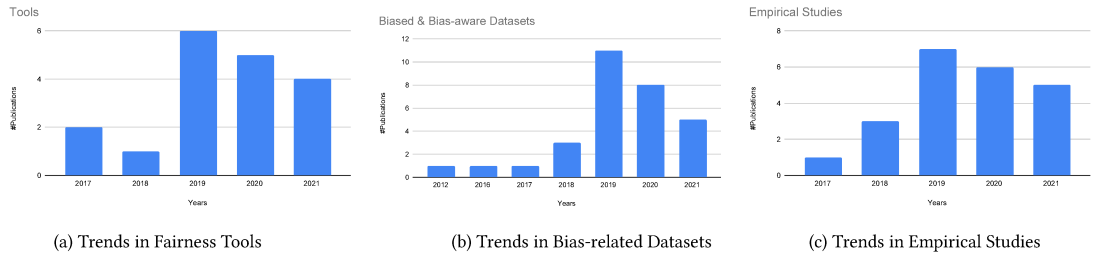

(a) Trends in Fairness Tools  (b) Trends in Bias-related Datasets  (c) Trends in Empirical Studies

Fig. 8. Details of trends in Fairness Tooling, Datasets and Empirical Studies

Table 2. Purpose of Software Fairness Analysis

| Categories | Sub-category | Description | #Pubs | Sample Works |
|---|---|---|---|---|
| **Validation** | Testing | generating discriminatory test inputs to expose fairness violations | 20 | [56, 130, 143, 144] |
| | Mitigation | mitigating bias in software systems, e.g., via repair and prevention | 14 | [8, 34, 57, 125] |
| | Debugging | diagnosis and explanation of fairness violations | 8 | [35, 88, 121] |
| | Auditing | analsying and measuring bias in software systems | 2 | [28, 82] |
| **Verification** | Verifiers | verifying that a system fulfills a fairness metric or goal | 12 | [7, 14, 59, 70] |
| | Certification | certifying that a system fulfills a fairness goal | 4 | [49, 116] |
| | Proof Guarantees | providing a formal proof that a system achieves a fairness goal | 2 | [32, 99] |
| **Design** | Requirements | requirement engineering and formalization of fairness properties | 4 | [17, 52, 90] |
| | Bias-aware Design | designing fair systems and bias-aware software | 15 | [29, 67, 74] |
| **Empirical Evaluation** | Analysis | empirical studies about fairness concerns | 22 | [21, 25, 41, 147] |
| | Benchmarking | providing fair benchmarks or benchmarks for fairness evaluations | 4 | [20, 24, 65, 133] |
| **Datasets** | Bias in Datasets | studying biases in training and evaluation datasets | 30 | [30, 58, 132, 138] |
| | Bias-aware Datasets | developing unbiased or bias-aware datasets for better evaluation | 5 | [25, 108, 113, 114] |
| **Tooling** | Automatic | providing fully automatic tools for fairness analysis | 18 | [7, 15, 19, 59] |
| | Semi-automatic | building tools that require human interaction for fairness analysis | 2 | [28, 89] |

### 5.2 RQ2 Purpose of Fairness Analysis.

In this study, we investigate the purpose of each paper collected in this survey, particularly, we examine and categorise the fairness problem studied or addressed by each paper. Table 2 provides details of the purpose of software fairness analysis performed in the literature. Overall, we identified six (6) categories for all collected papers. In the following, we discuss the problem addressed in each category, and the notable works that address such issues. In addition, we highlight the gaps in each category and across all identified categories. Figure 9 and Figure 10 show the the research focus of each paper and the purpose of the fairness analysis conducted in each of these papers, respectively. We also examine how the research community analyzes software fairness, especially the SE, PL and Security venues. We are

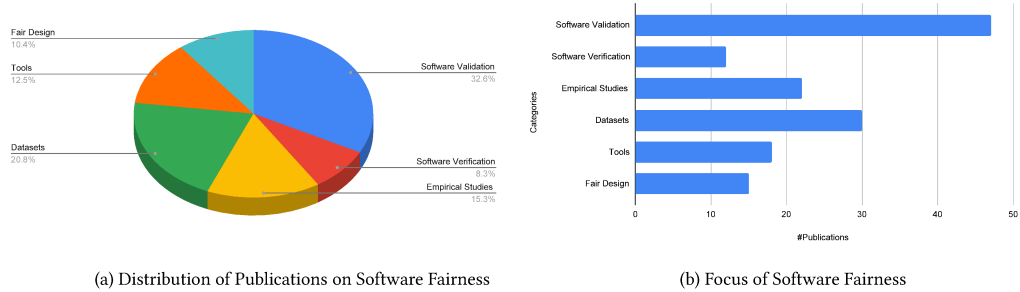

(a) Distribution of Publications on Software Fairness  (b) Focus of Software Fairness

Fig. 9. General Focus of Fairness Analysis

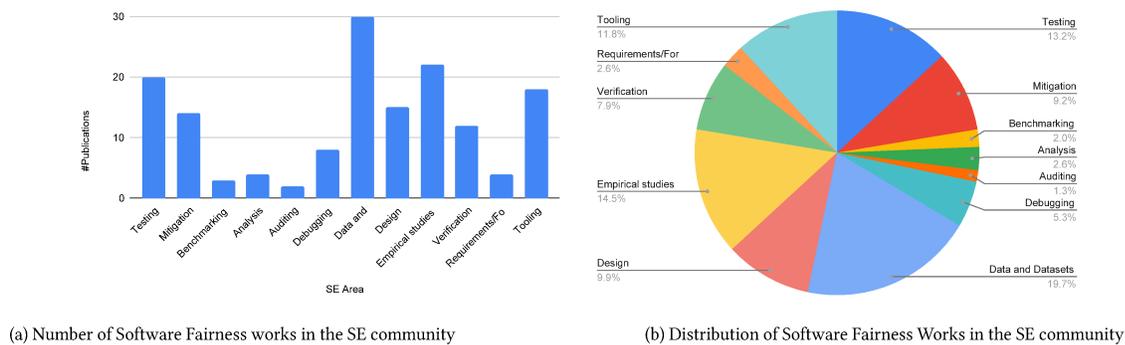

(a) Number of Software Fairness works in the SE community  (b) Distribution of Software Fairness Works in the SE community

Fig. 10. Purpose of Fairness Analysis in SE community

interested in the focus of the community, the areas that are well investigated by the community, and the areas that are not explored. Table 3 provides high level details of some of these publications.

Generally, we observed that *the focus of the research community has been on the validation, design and empirical studies of software fairness.* Particularly, one in three (about 33% of) the collected papers study fairness validation, i.e, the testing, debugging and mitigation of fairness errors (*see* Figure 9(a)). Analogously, more than one-third of collected papers (together over 36%) either study biases in datasets (about 21%) or conduct empirical evaluation of software fairness (about 15.3%). Figure 10 shows the focus of the papers published in SE venues. An inspection of these works further show that *dataset analysis, empirical studies, testing and tooling of software fairness is popular* in the community (*see* Figure 10 (a)). In particular, fairness *debugging, requirements analysis, benchmarking, and auditing are not popularly studied in the community* (*see* Figure 10(b)). On one hand, this suggests that the focus of the majority (70%) of the (SE) research work is focused on the validation, empirical evaluation and data(sets) analysis of fairness properties. On the other hand, concerns such as the auditing, debugging and requirements analysis are under-studied.

> *Most publications (70%) study the validation, empirical evaluation and data(set) analysis of software fairness: The design, verification and tooling of fairness-aware software have been under-studied (about 30%).*

***Bias Testing, Debugging and Mitigation.*** The research papers in this category aim to detect, expose, diagnose or mitigate fairness issues in learning-based software systems. Figure 10 provides the distribution for each category. The bulk of this work are general-purpose approaches proposed in the SE, PL and security venues, while the rest of the

Table 3. Excerpt of works performing Software Fairness Analysis ("Acc." means "level of software access")

| Acc. | Approach | Goal | Problem | Main Idea | Core Technique |
|---|---|---|---|---|---|
| Black-box | LTDD [88] | Debugging | identifying biased features in training data | debugging biased features to build fair ML software. | data debugging, linear-regression |
| | Multiacc. Boost [82] | Auditing, Analysis, Mitigation | audit/mitigate multiaccuracy, i.e., group fairness for all subgroups | perform multiaccuracy audit and post-process models to achieve it | multiaccuracy auditing, post-processing |
| | Cito et al. [35] | Debugging, Mitigation | debug and isolate the cause of mispredictions in ML models | characterize the data on which the model performs poorly | rule induction |
| | ASTR-AEA [121] | Debugging, Testing, Mitigation | performing fairness testing without existing datasets | discover and diagnose fairness violations in NLP software | grammar-based testing |
| | Fair-Way [34] | Debugging, Testing, Mitigation | detect and explain how ML model acquires bias from training data | identify how ground truth bias affects ML fairness | multi-objective optimization, pre/in -processing |
| | FairSM-OTE [33] | Debugging, Testing, Mitigation | finding biased labels in training data generation | remove biased labels and balance data using sensitive attribute | situation testing, data balancing |
| | Fair-Vis [28] | Auditing, Analysis | auditing and analysing group fairness in ML model | visual analytics for the discovery and audit of (sub)group fairness | visual analytics, domain knowledge |
| | Flip-test [22] | Testing, Mitigation | testing individual fairness – similar treatment of protected statuses | discover individual (un)fairness and its associated features | optimal transport, flipset, distr. sampling |
| | Aequitas [130] | Testing, Mitigation | validation of fairness for arbitrary ML models? | generating discriminatory inputs to uncover fairness violations | directed testing, probabilistic search |
| | Themis [10, 26, 56] | Formaliz., Testing | formalize software fairness testing for casual discovering of discrimination | measure causal discrimination in software to direct fairness testing | input schema, causal relationships |
| | Aeque-Vox [105] | Debugging, Testing | testing group fairness for Automatic Speech Recognition (ASR) systems | group fairness testing by simulating different environments | ML robustness, test simulation, fault localization |
| | ExpGA [47] | Testing | current individual fairness testing methods suffer poor efficiency, effectiveness, and model specificity | fairness testing by modifying feature values using explanation results and genetic algorithm (GA) | genetic algorithm, feature mutation, search based testing |
| | CGFT [100] | Testing | Uneven distribution of fairness tests and variations in execution results | leverage combinatorial testing to generate evenly-distributed test suites | combinatorial testing, input coverage |
| | SG [5] | Testing | detecting the presence of individual discrimination in ML models. | auto-generation of test inputs for detecting individual discrimination. | symbolic execution, local explainability |
| | Bias-Finder [11] | Testing | Bias testers for SA systems rely on small, short, predefined templates | discover biased predictions in SA systems via metamorphic testing. | template curation, NLP techniques, metamorphic testing |
| | Biswas and Rajan [20] | Mitigation | understanding fairness characteristics in ML models from practice | empirical evaluation of fairness and mitigations on real-world ML models | empirical study |
| | Fairea [65] | Mitigation | what is the SE trade-off between accuracy and fairness? | benchmarking and quantifying the fairness-accuracy trade-off achieved by bias mitigation methods | model behaviour mutation |
| White-box | ADF [144, 145] | Testing | searching individual discriminatory instances | generating discriminatory inputs violating individual fairness via ML | gradient computation and clustering |
| | Deep-Inspect [127] | Testing | detecting confusion and bias errors at class-level | expose confusion and bias errors in image classifiers | class property violations, robustness |
| | EIDIG [143] | Testing, Mitigation | how to detect and improve individual fairness of a model | generating test cases that violate individual fairness | gradient descent, global/local search |
| | Neuron-Fair [148] | Testing, Mitigation, Analysis | interpretability, performance, and generalizability in bias testing | identifying biased neurons, i.e., neurons that cause discrimination | neuron activation, adversarial attacks |
| | Fair-Neuron [57] | Mitigation, Analysis | balancing accuracy-fairness trade-off without additional model(s) | detect neurons with contradictory optimization directions, and achieve trade-off via selective dropout | joint-optimization, adversarial game |
| Grey | Tizpaz-Niari [128] | Debugging, Testing, Mitigation | explaining fairness impact of hyper-parameters | identify the effect of parameters on software fairness | search based testing, statistical debugging |
| | CAT/TransRepair [125, 126] | Testing, Mitigation | detecting inconsistency in machine translation (MT) | detect inconsistency bugs without access to human oracles | mutation testing, metamorphic testing, language model (BERT) |

proposed approaches are more *specialised* approaches proposed for analysing applications in specific domains, e.g., CV

or NLP venues. Overall, we found that most approaches are focused on fairness testing, some methods are focused on the mitigation of unfairness, while very few techniques are focused on debugging (diagnosing and understanding the root causes of) unfairness. In the following, we shed more light on the main idea of available approaches and the gaps in techniques.

*Fairness testing* techniques employ a plethora of techniques for test generation, including ML, search and program analysis techniques. Table 3 provides details on the fairness testing methods proposed in the literature. These test generation methods include several black box testing approaches (especially, input-based approaches), and few white-box techniques. Notably, *there are even fewer grey-box fairness testing approach*. Thus, most proposed approaches drive the generation of discriminatory inputs either by *analysing the model under test (MUT)* or the *input space*. However, there are very few techniques that leverage both the input space and the model analysis, besides, there are few studies investigating the relationships between both dimensions for testing purposes (e.g., Tizpaz-Niari et al. [128]). Notably, white box approaches (e.g., ADF [144, 145] and EIDIG [143]) mostly employ ML techniques (e.g., gradient computation, and clustering) to drive the generation of discriminatory test cases. Meanwhile, black-box approaches focus on leveraging the knowledge of the input space, program analysis and/or search algorithms to generate discriminatory inputs. They mostly employ templates, schemas, grammar, mutation or search algorithms to drive fairness test generation [121, 125, 130, 137]. Other approaches employ program analysis, e.g., symbolic execution [5] and combinatorial testing [100] to drive the generation of discriminatory test inputs. Notably, *we found few grey-box testing techniques that leverage both the input space and internal model attributes/properties to drive the generation of discriminatory inputs*. Moreover, there is little work that studies the link between the properties of the input space or discriminatory inputs to other internal model attributes/properties for fairness testing (e.g., Tizpaz-Niari et al. [128])).

> *Fairness testing approaches are mostly black or white box test generation methods: There are few grey-box approaches that leverage (or study) the relationship between the input space and internal model properties.*

**Fairness Verification, Certification and Proof Guarantees:** We examine the literature on the verification, certification and proving of fairness properties in learning-based software systems. Specifically, we examine the categories of verifiers, the main ideas of proposed verification approaches and the gaps in this research area. To this end, we identify three major kinds of fairness verification approaches, namely *distributional verifiers* (e.g., FairSquare [7] and VeriFair [14]), *specialised verifiers* designed for a particular domain, metric or task (e.g., [32, 49, 70, 91, 95]) and sample-based verifiers (e.g., AIF360 [15] and Themis [56]). In our study, most approaches are *distributional-based* verifiers or *specialised* verification techniques, and there are fewer *sample-based verifiers*. Distributional verifiers typically encode fairness metrics as probabilistic properties then verify such properties with respect to the underlying data distribution of the learning-based system. On the other hand, sample-based approaches (e.g, AIF360) generate tests to verify fairness metrics based on a fixed test dataset, otherwise they generate counter-examples to refute the satisfaction of the fairness property. Other verification approaches target specific domains (e.g. NLP [95] ), specific fairness metrics (e.g., individual fairness [70] and disparate impact [49]), or tasks (fair training [32] and data debiasing [49]).

Some verification approaches encode fairness metrics as probabilistic properties, then provide guarantees over the system's data distribution. These approaches take as input the probability distribution of the attributes in the dataset and the MUT, then verify the fairness of the system with respect to the distribution and MUT. For instance, FairSquare [7] presents a technique for verifying and certifying fairness properties by encoding fairness definitions as probabilistic properties. Moreover, Albarghouthi et al. [8] also proposed an approach that applies *distribution-guided inductive synthesis* to verify (and repair) *unfair* ML classifiers. Likewise, Bastani et al. [14] developed an algorithm for verifying fairness specifications (called VeriFair), the algorithm provides probabilistic guarantees for fairness properties and allows users to verify that the probability of fairness errors is small. Ghosh et al. [59] presents a stochastic satisfiability (SSAT) framework (called Justicia) to formally verify fairness measures of supervised learning algorithms with respect to the underlying data distribution. Justicia is applicable to different fairness metrics including disparate impact, statistical parity, and equalized odds. Compared to previous distribution-based verification approaches (FairSquare [7] and VeriFair [14]), Justicia supports non-Boolean and compound sensitive attribute. It also provides theoretical bound for the finite-sample error of the verified fairness measure, and it is more robust than sample-based verifiers (e.g., AIF360).

There are several verification approaches that are focused on a specific domain, metric, or task (e.g., debiasing datasets [49] or fair training [32]). For domain- or task-specific approaches, Ma et al. [95] provides a black-box technique to enforce fairness guarantee for NLP systems by leveraging advances in certified robustness of machine learning. Their approach employs a neutral phase to piggyback the NLP model to smooth its outputs such that they are certified to preserve individual fairness. Similarly, Liu et al. [91] presents a (semi-)automated verification framework (called FairCon) to ascertain the fairness of smart contracts, their approach can refute false claims with concrete examples, or certify that contract implementation fulfil desired fairness properties.

For metric-specific approaches, John et al. [70] proposes sound (but incomplete) verifiers for proving individual fairness of models by employing appropriate relaxations of the problem, specifically for linear classifiers and kernelized polynomial/radial basis function classifiers. Likewise, Feldman et al. [49] presents a verification technique for certifying the (im)possibility of disparate impact on a data set by employing a regression algorithm that minimizes the balanced error rate (BER) of the dataset. The goal is to verify that a protected or sensitive attribute can not be predicted from the other attributes in the dataset by ascertaining if there is sufficient information about the dataset to detect sensitive attributes from the data. In terms of verifying fair training, Celis et al. [32] propose a technique to train fair classifiers with theoretical guarantees, using a meta-algorithm for classification that can take as input a general class of fairness constraints with respect to multiple non-disjoint and multi-valued sensitive attributes. Notably, this approach can handle non-convex fairness constraints such as predictive parity.

A different line of fairness verification techniques are sample-based verifiers such as AIF360 [15] and Themis [56]. Typically, these approaches leverage software testing techniques to verify fairness properties on fixed data sample. AIF360 [15] is an extensible open source toolkit for testing and verifying fairness properties, particularly for a fixed data sample. It provides an array of testing methods to generate discriminatory test suites, and report several fairness performance metrics for any fairness algorithm. It also allows to perform both unit and integration testing for bias mitigation algorithms. Meanwhile, Themis [56] allows developers to verify that a fixed sample do not discriminate against specific sensitive attributes (e.g., race and gender) by automatically generating discriminatory that verifies that changing the instance of the attribute does not cause a change in the output of the learning-based system. Overall, these approaches leverage advances in software testing to measure and verify that a fixed sample data fulfills specific fairness properties.

There are other general verification approaches that verify multiple (user-defined) fairness constraints or other properties beyond, but including, fairness properties. As an example, Metevier et al. [99] addressed the problem of verifying multiple fairness definitions as well as user-defined fairness metrics for learning-based systems. The authors proposed a verification approach (called RobinHood) which employs an offline contextual bandit algorithm determine the satisfiability of several fairness constraints. Morevoer, RobinHood provides a probabilisitic guarantee of fairness by ensuring that it does not return a solution with a probability greater than a user-defined threshold. Besides, Sharma et al. [117] is a more general verification approach which is also applicable to fairness properties. The authors propose a (white-box) verification approach that employs the knowledge of the internal structure of the model to verify that ML models fulfil several properties (including fairness), in particular by training a shadow model that approximates the MUT by using the prediction of the original model as training data. It employs a property specification language to test and verify model properties of learning-based software and provides counter-examples (i.e., test cases violating the property) if the property is not fulfilled.

> *Most fairness verifiers are distribution-based, sample-based or specialised for a specific fairness measure, domain or task. There are very few verifiers that support multiple or user-defined fairness constraints.*

**Others:** Our investigation showed that the SE research community has *mostly* focused on analysing software fairness as a *fairness validation (i.e., testing and debugging)* problem [56, 130] and as a *fair system design* problem, especially to mitigate biases [27, 34, 129]. However, other aspects of SE concerns are under-studied, such as the formalization of fairness as a *software requirement* [26, 52], the *verification* of fairness properties [7], and *empirical evaluation* of software fairness properties [20, 65]. Furthermore, some aspects of SE concerns are hardly studied by the community, namely, empirical evaluation of fairness properties (especially human factors in software fairness [61]) and the maintenance of fairness properties as the software evolves (e.g., because of model re-training, model compression [64] or software regression).

In 2017, Brun and Meliou [26] formalised software fairness as a software engineering problem that needs to be tackled by all aspects of software engineering. These aspects include steps in the software development life cycle such as requirements engineering, design, testing, verification and maintenance. Since their publication, the bulk of the published papers have *focused on the design, testing and mitigation* of software fairness, these areas have been well explored and investigated by these communities. For instance, several papers have explored the problem of *fairness testing* by employing random test generation (Themis) [10], local search algorithms (Aequitas) [130], gradient computation (ADF and EIDIG) [143–145], mutation testing (TransRepair) [125], grammar-based testing (Astraea) [121], symbolic execution [5], property-driven testing [117] and schema or template based testing (BiasRV) [137]. In addition, the problem of fair system design to mitigate biases has been studied by a few researchers via several techniques including behavior mutation [65], and feature or dataset manipulation [141].

Very few researchers have conducted empirical evaluation of software fairness properties in real-world applications. Notably, Biswas and Rajan [20] conducted an empirical study to study several fairness mitigation techniques including their impact on performance. Likewise, Hort et al. [65] conducted a large scale empirical study to test the effectiveness of 12 widely-studied bias mitigation methods. Meanwhile, Zhang and Harman [141] empirically evaluated how feature set and training data affect fairness. The focus of most studies has been on fairness mitigation, except Zhang and Harman [141] that empirically studied the impact of features and datasets on fairness properties. Besides these two concerns, we have found very few studies in these fields empirically evaluating other SE concerns (such as human factors [38, 63]) or concerns relevant to steps of the model/software development pipeline.

However, some aspects of software fairness analysis remains under-investigated. Firstly, there is little work addressing concerns about the maintenance of software fairness, for instance, as the software or model evolves over time (i.e., software regression analysis) or is optimized (e.g., via model compression for edge devices). For instance, although there is a recent paper investigating the impact of model compression on software fairness [64], we found no work to support the software engineering activities around such changes, such as testing for such changes, e.g., in a regression scenario. Likewise, there are very few empirical studies on software fairness properties, particularly, there has been few empirical evaluation in this area involving humans, which is vital to determine the harm caused by unfair software behavior [38, 63]. Secondly, the formalization and definition of fairness as a software requirement, metric or measure has only been performed by a few researchers, we found few (about three) papers in this area, namely Brun and Meliou [26], Verma and Rubin [131] and Finkelstein et al. [52]. We aslo found only few papers (e.g., FairSquare [7]) in the area of fairness verification. Other verification approaches, e.g., Finkelstein et al. [52], tackled the verification of fairness properties by encoding fairness definitions as probabilistic program properties, then automatically verifying and certifying that a program meets a given fairness property. We encourage that more effort has to be invested in verifying, certifying and providing proof guarantees of fairness properties in learning-based software systems.

> *The SE research community mostly study software fairness properties as mitigation, design and testing problems. Very few works have studied fairness as a requirements engineering or verification problem, and fewer works empirically studying human factors of fairness properties and how to maintain fairness as software changes/evolves.*

### 5.3 RQ3 Fairness measure.

In this research question (RQ3), we examine the fairness metrics analyzed by the studies and methods proposed for software fairness analysis. We investigate the number of studies examining the different classes of fairness metrics (e.g., statistical measures, similarity-based measures and time-based measures), as well as the distribution of specific metrics, such as individual fairness, group fairness and causal fairness. Figure 11 highlights our analysis of the distribution of these metrics across proposed methods and conducted studies.

Generally, we observed that most studies examine statistical or similarity based measures (such as individual, group or causal fairness), while time-based metrics (e.g., sequential or long-term fairness) or measures based on causal reasoning (e.g., fair inference) are not (yet) studied in the SE community as software fairness metrics. Statistical and similarity based fairness metrics (such as group, individual, and intersectional fairness) are the most examined metrics in software engineering (see Figure 11(a)). For instance, individual fairness and group fairness account for more than three in four (76.5%) of studies and methods in software fairness analysis. Causal reasoning based fairness metrics (such as causal fairness) are also commonly studied. However, time-based metrics were not found in the SE literature [146].

> *Statistical and similarity -based fairness metrics are popularly studied (86.5%) in fairness analysis, but metrics hinged on causal reasoning are under-studied (about 10%) and we could not a find a single work studying time-based metrics.*

In addition, we observed that fairness metrics such as *individual and group fairness metrics are well studied in the SE community*. Concretely, about 42.5% and 35% of publications in software engineering venues study individual fairness and group fairness, respectively. The focus of most of these studies is in terms of testing, validating and mitigating software to fulfil these metrics. Several of these techniques are tailored towards testing, discovering and mitigating one or both of these metrics. Concretely, 27% of examined studies study only individual discrimination [130, 143, 144]. For instance, Zhang et al. [143] proposed EIDIG (Efficient Individual Discriminatory Instances Generator), a scalable and efficient approach to systematically generate test cases that violate the individual fairness for DNN models. Likewise, Zhang et al. [144] proposed approaches to search for individual discriminatory instances of DNN, using lightweight procedures like gradient computation and clustering. Overall, we found that 23% of examined papers in the community study both individual and group fairness, 17% examine only group fairness and 27% study only individual fairness.

In our analysis, causal fairness is also a commonly studied fairness measure in the community, it accounts for one in ten publications. Notably, Galhotra et al. [56] and Biswas and Rajan [21] have studied the testing and debugging of

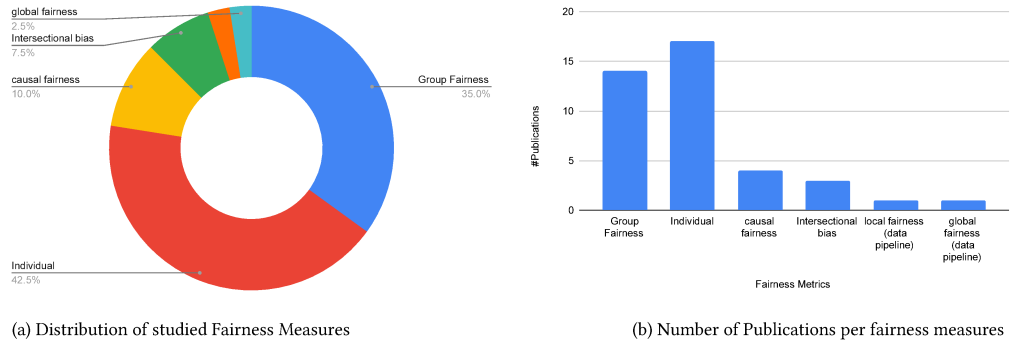

(a) Distribution of studied Fairness Measures         (b) Number of Publications per fairness measures

Fig. 11. Details of Fairness Measures

causal fairness. Meanwhile, other metric such as intersectional bias, local and group fairness (especially in ML or data piplines) are the least studied in the community, with each accounting for less 2.5 to 7.5 percent of all examined works. As an example, only Chakraborty et al. [33], [82] and Cabrera et al. [28] studied the analysis of intersectional bias in SE venues, specifically, in terms of testing, auditing and visual analysis of intersectional bias, respectively. There is still a lot of work to do in this area, particularly, how to effectively test, mitigate and debug intersectional bias in software.

Finally, On the extreme end, we found almost no paper in the SE community studying time-based fairness metrics such as sequential and long-term fairness [146]. These metrics are important because of maintaining fairness properties as the software evolves. As an example, consider an automated classifiers who is re-trained periodically dues to new data requirements, how do we ensure that such data and the resulting classifier maintains the fairness property defined in the previous system? Overall, this shows that the focus of the community is on a set of statistical and similarity-based fairness metrics (such as individual, group, causal, hence, ignoring other metrics, e.g., time related bias concerns.

> *Most SE studies (86.5%) investigate individual fairness, group fairness and causal fairness metrics, metrics such as intersectional fairness are under-studied, and time-based metrics such as sequential and long-term fairness have not been studied in the SE community.*

## 5.4 RQ4 Bias and Sensitive Attributes.

Let us investigate the societal biases studied in software fairness. In particular, we analyze the sensitive or protected attributes (e.g., age, race, gender etc) that researchers study or develop techniques for. Figure 11 and Figure 12 illustrates the distribution of sensitive attributes in the literature. In Table 4, we exemplify and illustrate some of these sensitive attributes.

We observed that *about four in every five studies examine four main protected attributes, namely age, race, gender and country*. These four attributes are studied in about 79.5% of works, with age, race and gender accounting for the majority (about 72.5% of all works). We believe this is due to the availability of these attributes in many structured datasets, and the ease of manipulation. Notably, these protected attributes are common in structured datasets (e.g., tabular data) and they are easier to manipulate, test, and mitigate for structured datasets. For instance, attributes such as age, gender and race are popular features in structured datasets (e.g., German Credit, Adult Census Income and Bank Marketing). These attributes are popularly studied due to the simplicity of manipulating them in the employed structured datasets.

Table 4. Details of studied biases (in SE communities), i.e., sensitive/protective attributes. Examples are based on a simple sentiment analysis (SA) system that detects the emotional situation or state in a sentence, ☺ indicates a positive sentiment (e.g., happy),☹ indicates a negative sentiment (e.g., sad), bias-inducing inputs are <u>underlined</u> and unfair outputs are in <span style="color:red">red</span> (e.g. ☹ ).

| Sensitive Attributes | #Pubs | Illustrative Example for text-based SA system | Sample Works |
|---|---|---|---|
| *Gender* | 24 | "The {man/<u>woman</u>} is happy." = ☺/☹ | [33, 34, 130] |
| *Race* | 21 | "The {white man/ <u>black man</u>} is happy." = ☺/☹ | [56, 144, 145] |
| *Age* | 15 | "The {young man/ <u>old man</u>} is happy." = ☺/☹ | [20, 141, 143] |
| *Country* | 6 | "The {american man/ <u>chinese man</u>} is happy." = ☺/☹ | [105, 145] |
| *Language* | 2 | "The {english orator/ <u>spanish orator</u>} is happy." = ☺/☹ | [125, 126] |
| *Occupation* | 2 | "The {manager/ <u>cleaner</u>} is happy." = ☺/☹ | [11, 121] |
| *Religion* | 2 | "The {nun/ <u>atheist</u>} is happy." = ☺/☹ | [121, 145] |
| *Ethnicity/Accents* | 2 | "The {european man/ <u>asian man</u>} is happy." = ☺/☹ | [105, 145] |
| *Class label (e.g., poverty level)* | 1 | "The {rich man/ <u>poor man</u>} is happy." = ☺/☹ | [33, 127] |
| *Narcotics Arrests /Gang Affiliation* | 1 | "The {convict/ <u>innocent man</u>} is happy." = ☺/☹ | [22] |
| *Work experience* | 1 | "The {experienced farmer/ <u>novice farmer</u>} is happy." = ☺/☹ | [22] |
| *Academics (LSAT /GPA)* | 1 | "The {honors student/ failing student} is happy." = ☺/☹ | [22] |
| *Marital-status/Relationship* | 1 | "The {married man/ <u>single man</u>} is happy." = ☺/☹ | [28] |

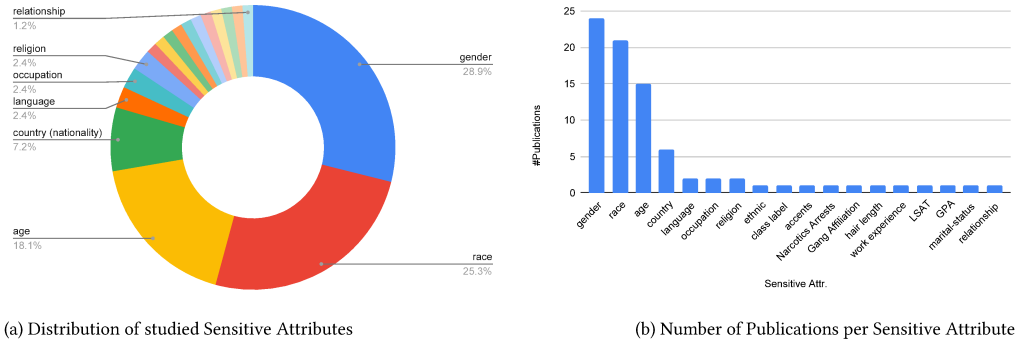

(a) Distribution of studied Sensitive Attributes     (b) Number of Publications per Sensitive Attribute

Fig. 12. Details of Biases, i.e., Sensitive/Protected attributes

They typically have an easily tractable input space, constraints and range. As an example, it is easy to manipulate gender features in tabular date, since there have a small number of possible values (e.g., male, female or non-binary). Meanwhile, rare protected attributes (e.g. hair length or narcotics arrest, accents and language features) either possess a complex input space or are more common in unstructured datasets (e.g., text, audio and images). Hence, they are more difficult to study or manipulate. For instance, consider an image dataset where the gender-related sensitive attribute are related to pixel values. This analysis suggests that the focus of the community has been on the sensitive attributes that are available in less complex tasks or structured datasets. This is evident by the fact that these attributes are commonly provided in most datasets employed in the community, particularly, structured datasets (see RQ5).

> *About four in five (79.5%) works study age, gender or race as sensitive attributes, while model class-label and specific attributes such as relationship (status), class (academics, work) and religion are understudied (less than 2.5% each).*

### 5.5 RQ5 Datasets and Tasks.

For this research question, we examine the datasets examined or employed in the collected publications. Table 5 provides details of the tasks and datasets employed in (the evaluation of) software fairness analysis. Furthermore, Figure 13,

Table 5. Details of Tasks and Datasets employed in Software Fairness Analysis

| Type | Task Category | #Data. | #Pubs | #Tasks | Tasks (Example Pubs) | Datasets (#Pubs) |
|---|---|---|---|---|---|---|
| Structured | Education | 2 | 3 | 1 | academic performance [22] | Law School (1) |
| | Finance | 6 | 54 | 4 | income [5, 141, 144], | Adult Census (19) |
| | | | | | credit default [10, 21, 34] | German Credit (17), Default Credit (3), Home Credit (3) |
| | | | | | potential buyers [21, 33, 88] | Bank Marketing (11) |
| | | | | | fraud [5] | Fraud Detection (1) |
| | Legal | 3 | 10 | 3 | recidivism [28, 34, 65] | COMPAS (8) |
| | | | | | arrests [22] | Chicago Strategic Subject List (SSL) (1) |
| | | | | | US executions [5] | US Executions (1) |
| | Medical | 5 | 11 | 5 | medical expenditure [33, 88, 148] | MEPS (5) |
| | | | | | heart disease [33, 34] | Heart Health (3) |
| | | | | | heart failure [35] | Heart Failure (1) |
| | | | | | cancer risk [35] | Cervical Cancer (1) |
| | Other (HR) | 1 | 1 | 1 | hiring [22] | Lipton (1) |
| | Other (rentals) | 1 | 1 | 1 | car rentals [5] | Raw Car Rentals (1) |
| | Other (shipwreck) | 1 | 2 | 1 | shipwreck survival | Titanic ML (2) |
| Unstructured | CV (image) | 7 | 10 | 2 | face detection | ClbA-IN (1), PPB (1), LFW (2) |
| | | | | | image recognition | COCO (2), imSitu (1), CIFAR (2), ImageNet (1) |
| | NLP (text, speech) | 12 | 10 | 6 | SA [11], CoRef [121], MLM [121] | Twitter (1), IMDB (1), EEC Dataset (1), Labor statistics (1) |
| | | | | | toxicity | Wiki Comment (1), Jigsaw Comments (1) |
| | | | | | machine translation | News Commentary (2) |
| | | | | | ASR [105] | Speech Accent Archive (1), RAVDESS (1), Multi speaker Corpora of the English Accents in the British Isles (1), Nigerian English speech dataset (1) |
| | SE (code) | 4 | 4 | 4 | Programs [35] | Bug2Commit (1), Diff Review (1), Code AutoComplete (1), Oncall Recommendation (1) |

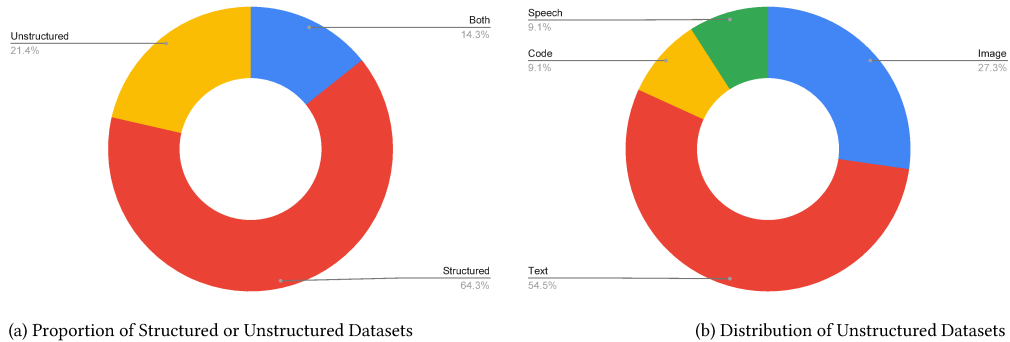

(a) Proportion of Structured or Unstructured Datasets  (b) Distribution of Unstructured Datasets

Fig. 13. Type of Studied Datasets

Figure 14 and Figure 17 highlight the type of the dataset (e.g., structured, or unstructured), the distribution of examined datasets, and the task associated with each dataset, respectively. We also examine the task category associated with the examined datasets (e.g., NLP, CV, etc), the volume of publications associated with each task category, and dataset (*see* Figure 15, Figure 18 and Figure 19, respectively).

***Volume of Publications per dataset:*** We found that *more than half of the examined studies employ four (4) major datasets, most of which are simpler, structured datasets for finance-related tasks.* Figure 13 illustrates the distribution of publications using each dataset. Notably, the most common datasets are the Adult Census Income, German Credit, Bank Marketing and COMPAS dataset, they are employed in over half (52%) of all (SE) papers. These datasets account for most

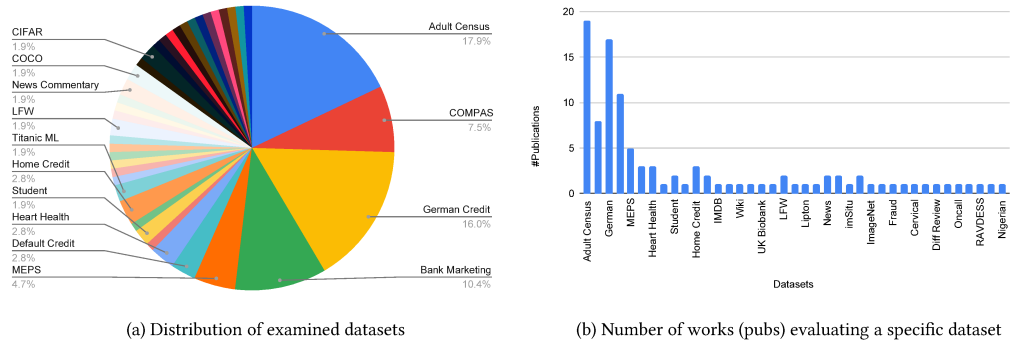

(a) Distribution of examined datasets  (b) Number of works (pubs) evaluating a specific dataset

Fig. 14. Details of examined Datasets

of the software fairness works we examined (*see* Figure 13 (b)). Most of these datasets are structured datasets for finance-related tasks such as income prediction, except COMPAS – a popular legal dataset for recidivism. However, structured datasets such as image (COCO, CIFAR) and text/NLP (News Commentary) are among the least studied datasets. This suggests that most research work employ similar datasets, which are mostly structured, tabular, less complex or finance related. Hence, implying there is the need for a more diverse evaluation of fairness analysis techniques that are general and cuts across several datasets and tasks.

> *Simpler, structured datasets (e.g., Adult Census Income) are the most employed datasets in software fairness analysis. Complex, unstructured datasets (e.g., image – COCO or CIFAR) are less popularly studied.*

**Type of examined datasets:** We observed that *the majority (64%) of the research works study software fairness for structured datasets*. Figure 13(a) shows that about two-third of the examined papers study fairness properties relating to structured datasets such as tabular datasets (e.g., Adult Income Census). *Fewer studies (about 21% of papers) solely analyse fairness properties for unstructured dataset* (e.g., text, image, speech or code). *Research works that generalise to both structured and unstructured datasets are even fewer (about 14%)*. This finding implies that most fairness analysis techniques are specialised for structured or unstructured datasets. This suggests the need to develop and empirically evaluate more general fairness analysis techniques that are (directly) applicable to both structured and unstructured datasets.

> *Very few (14% of) fairness analysis techniques generalise to both structured and unstructured dataset: Majority (86%) of techniques are specialised for either structured or unstructured datasets.*

**Unstructured datasets:** Figure 13(b) provides the distribution of works studying unstructured datasets. Inspecting works studying fairness properties in unstructured datasets, we observed that most (about 72%) are focused on text (i.e., NLP-related) datasets or image (i.e., CV) datasets. *Datasets relating to speech (i.e., audio) and code (i.e., programs) are the least studied in software fairness*, accounting for about 18% of the examined papers. Overall, this suggests that the SE research community has been focused on studying fairness properties as related to structured (relatively less complex) datasets, while more complex, unstructured datasets are under-explored. Moreover, fairness concerns relating to unstructured datasets speech (i.e., audio) and code (i.e., programs) remain under-studied.

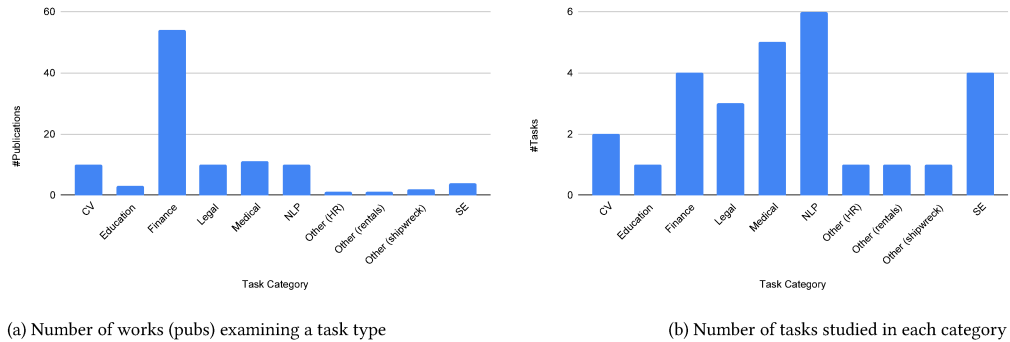

(a) Number of works (pubs) examining a task type

(b) Number of tasks studied in each category

Fig. 15. Details of Task Categories

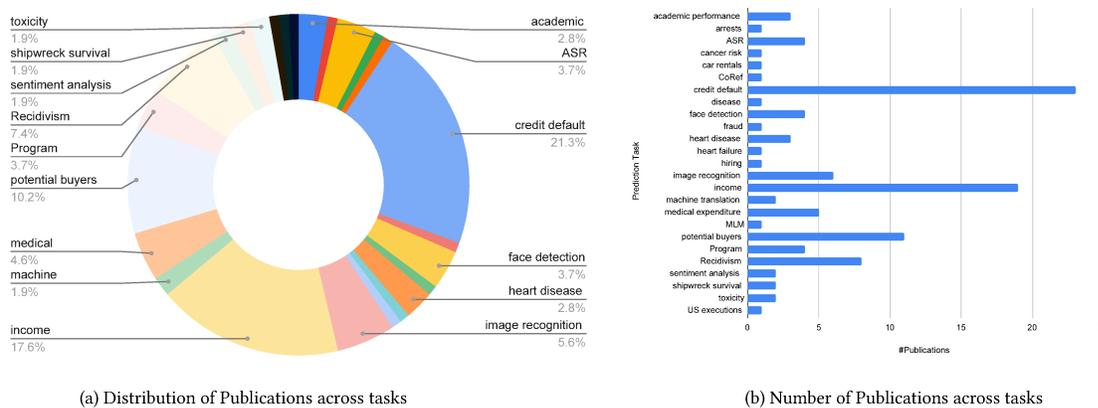

(a) Distribution of Publications across tasks

(b) Number of Publications across tasks

Fig. 16. Details of Publication across tasks

> *Fairness concerns relating to unstructured datasets such as code and speech (audio) are rarely studied, they account for only about 18% of all inspected papers.*

**Task Categories:** Figure 15 and Figure 17 highlight the details of the task categories and the datasets employed for each task (category). We found that *tasks involving finance, computer vision (CV) and natural language process (NLP) are the most studied in software fairness publications, both in terms of the number of tasks and the number of publications.* However, software fairness concerns for tasks involving *education* and *code analysis* are less frequently studied. Figure 15 illustrates that all of the well studied task categories (CV, NLP and finance) have up to two to four tasks each, while education had one task. As an example, finance-related publications contribute over 50 papers (*see Figure 15(a)*), with about four distinct tasks examined (Figure 15(b)). These tasks include predicting income, credit default, fraud and potential buyers (Figure 17). In addition, despite fewer publications, we observed that some task categories have more examined tasks. For instance, consider NLP task category, over six tasks have been studied in the literature, e.g., sentiment analysis, CoRef, MLM, ASR etc. This is despite very few publications for fairness analysis of NLP systems.

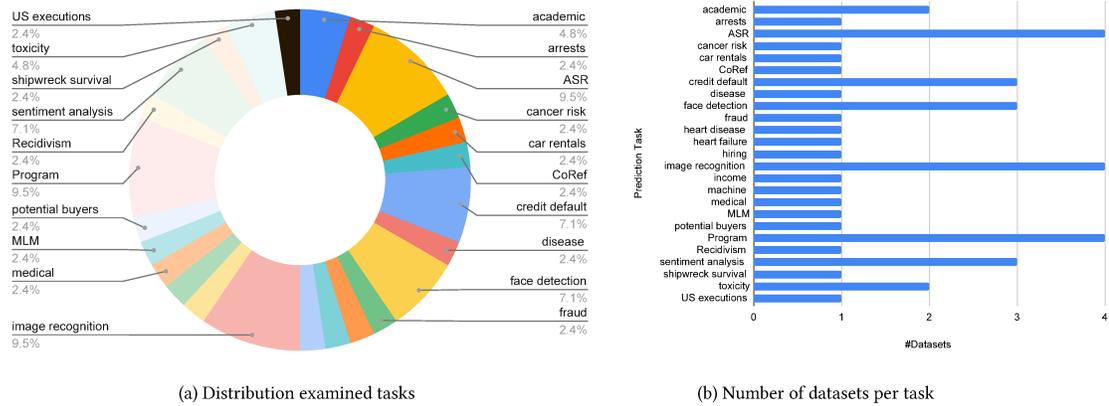

(a) Distribution examined tasks

(b) Number of datasets per task

Fig. 17. Details of Datasets examined for each Task

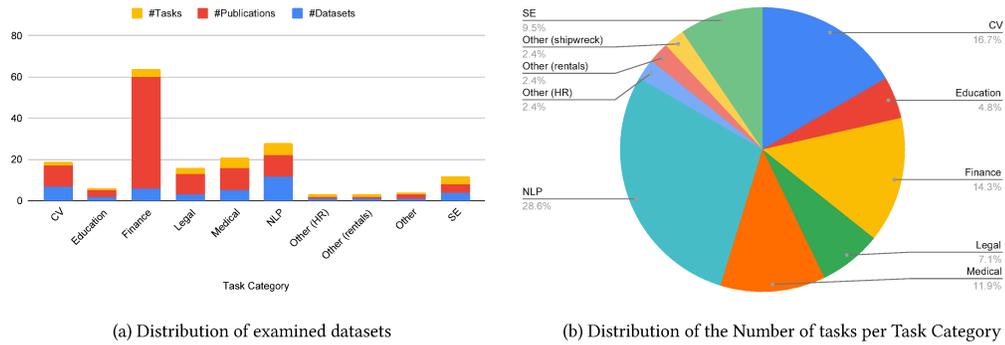

(a) Distribution of examined datasets

(b) Distribution of the Number of tasks per Task Category

Fig. 18. Details of Publications and Datasets for each Task category

Our analysis of the tasks investigated in the collected papers confirm that *financial, CV and NLP tasks are well studied for software fairness analysis*. Figure 18 and Figure 19 further shows the distribution of publications based on the number of tasks, and datasets. Similarly, Figure 16 shows that finance related tasks are the most studied, especially tasks such as credit prediction, income prediction and fraud detection. Figure 17 provides more fine grained analysis of the datasets relating to each task. Evidently, tasks involving finance, CV and NLP account for most examined datasets, about seven (7) to nine (9) percent each. For instance, tasks involving face detection, credit default, image recognition, speech recognition (ASR) and programs contribute about four (4) datasets each (*see* Figure 17(b)). These findings suggests that the community has been focused on investigating bias in specific sectors (CV, NLP and finance) while ignoring other categories (e.g., education and code).

> *The (SE) research community has been focused on studying software fairness concerns for (three) specific tasks (finance, CV and NLP tasks), but fairness concerns in areas like education and code analysis have been largely ignored.*

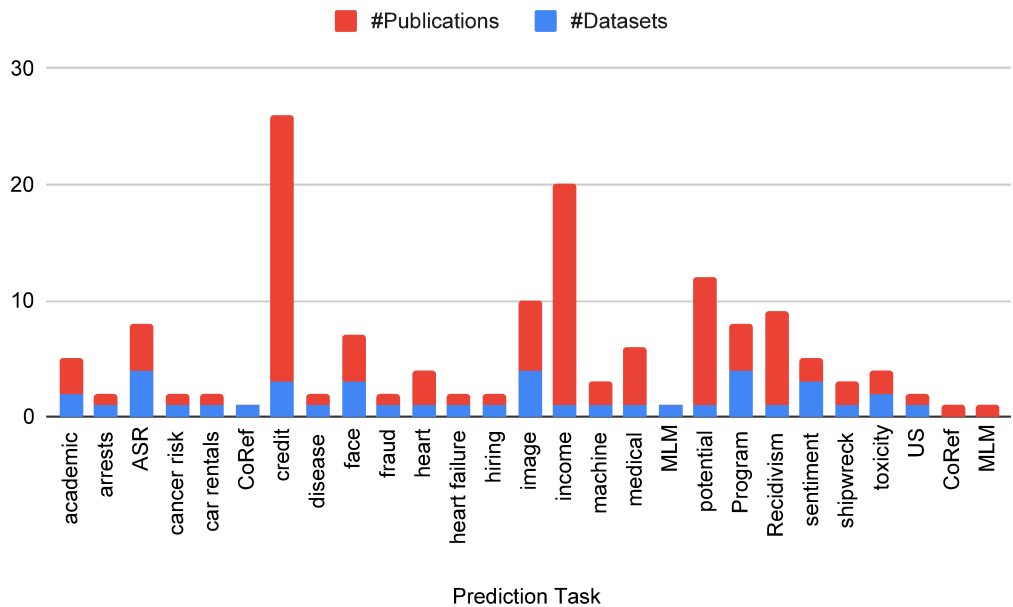

Fig. 19. Details of Publications and Datasets per Prediction Task

## 5.6 RQ6 Tooling

In this evaluation, we inspect the papers that propose a tool, framework or library to enable software fairness analysis. Table 6 provides details of some of the tools found in the literature. We categorize the goal of the analysis performed by each tool, the addressed problem, the main approach employed and their processing stage (pre, in and post -processing), as well as the software access required by the tool (black, grey or white box access).

Our analysis showed that *post-processing of AI software for validation purposes (e.g.. testing, auditing, analysis and mitigation) is the most prominent tool support available for software fairness analysis.* Table 6 shows that most tools are black-box post-validation tools. This is followed by support for fair learning (i.e., designing fair software systems). These includes measuring and analyzing the trade-off between fairness and accuracy metrics of the software, e.g., AIF360 [15] and POF [18]. Overall, *there is very little tool support for the specification, formalization and verification of fairness metrics.* Some tools support the verification of fairness properties in the post-processing stage, i.e., verifying trained model (e.g., FairSquare [7]). Meanwhile tools like VeriFair [14] verify fairness properties in the pre-processing stage, i.e., verifying if datasets fulfill a fairness property. There is also *little support for in-processing and white-box analysis of software fairness.* While there are some works that support all processing stages (i.e., pre, in and post) of software fairness [13, 15, 71, 120], we found low support for tools specifically focused on the in-processing stage.

Generally, fairness analysis tools are mostly automated, they provide a software architecture, API or framework which implements several (mitigation) algorithms that enable developers to conduct bias analysis. For instance, AIF360 [15] provides an extensible architecture, FAT-Forensics [120] offers a python framework providing several mitigation algorithms, and Themis-ML [13] provides an API for analysis. Interestingly, some approaches allow to visualise fairness metrics while auditing or analyzing fairness properties (e.g., Fairkit-learn [71]), and others enable fairness test experimentation. Notably, 2AFC [89] tests for (un)fairness via psycho-physical experimentation.

Table 6. Excerpt of Fairness Analysis Tools

| Tool (Paper) | Goal | Addressed Problem | Process. Stage | Approach | Access |
|---|---|---|---|---|---|
| Fairkit-learn [71] | Fair learning, Analysis | how to reason about and determine the trade-off between model quality (accuracy) and fairness | pre, & post | model search, visualisation | Grey |
| AIF360 [15] | Fair learning, Analysis | understanding how, when and why to use different bias handling algorithms in the model life-cycle | pre, in, & post | extensible architecture for analysing fairness metrics | Grey |
| POF [18] | Fair learning, Analysis | how to compute the "Pareto curve" of the trade-off between accuracy and fairness in the *regression* settings (*continuous* prediction/targeted values) | pre | fairness regularizers, Price of Fairness (PoF) metric | Grey |
| AITEST [6] | Testing | how to detect the presence of individual discrimination in ML models | post | symbolic execution and local explainability | Black |
| 2AFC [89] | Testing | how to relate unobservable phenomena deep inside models with observable, outside quantities that we can measure from inputs and outputs | post | Test Experiments, Experimental Psychology, Psycophysics, two-alternative forced choice (2AFC) | Black |
| Pc-fair-ness [135] | Formalization, Analysis | how to bound path-specific counterfactual fairness, address their *identifiability*, i.e., whether they can be uniquely measured from observational data | post | parameterized causal modelling, linear programming, response-function variables, constraints | Black |
| BiasRV [137] | Testing | how to monitor and uncover biased predictions at runtime | post | automatic template generation, mutation, metamorphic relations | Black |
| Themis [10] | Formalization, Testing | how to formally define and test software fairness using a causality-based measure of discrimination | post | causal inference, schema-based test generation | Black |
| Fair-Square [7] | Formalization, Verification, Certification | how to verify or certify that a program meets a given fairness property | post | probabilistic reasoning, SMT solving, symbolic weighted -volume-computation algorithm | Black |
| FAT Forensics [120] | Analysis, Auditing, certifying | how to inspect datasets (features), models and their prediction for fairness metrics | pre, in, & post | an inter-operable Python framework for fairness (FAT) algorithms | White, Black, & Grey |
| VeriFair [14] | Verification, Specification | how to verify fairness specifications, i.e., fairness properties of ML programs | post | adaptive concentration inequalities | Black |
| Justicia [59] | Verification | how to formally verify the fairness metrics are satisfied by different algorithms on different datasets | pre | stochastic satisfiability (SSAT) | Black |
| Checklist [107] | Testing | how to test (fairness) behaviors of NLP systems | post | behavioral testing, template-based test generation | Black |
| FairML [2] | Auditing, Analysis | how to determine the significance of inputs in assessing the fairness of black-box models | post | model compression, input ranking algorithms | Black |
| MT-NLP [95] | Testing, Mitigation | how to determine if NLP models are free of unfair bias toward certain sub-populations/groups | post | metamorphic testing | Black |
| ASTRAEA [121] | Debugging, Testing, Mitigation | how to perform fairness testing without an existing dataset, i.e., no training data access | post | grammar-based testing, metamorphic relations | Black |
| Aequitas [110] | Auditing, Analysis | how to audit for bias and fairness when developing and deploying algorithmic decision making systems | post | bias audit toolkit to support many bias metrics | Black |
| FairTest [129] | Testing, Debugging | how to detect unwarranted associations (UA) (disparate impact, offensive labels, and uneven error rates) between model outcomes and data attributes | post | unwarranted associations (UA) framework to determine UA between outcomes and attributes | Black |
| Themis-ml [13] | Fair Learning, Mitigation, Analysis, Auditing | how to measure, understand, and mitigate the implicit historical biases in socially sensitive data | pre, in & post | API for Fair ML for simple binary classifier | White, Black, & Grey |

> *Fairness Analysis tools are mostly focused on the post validation of AI-based software, with little (grey) or no access (black) to the AI model. There is a need for tools that support white-box, in-processing stages of fairness analysis, especially for specification, formalization and verification.*

## 6 LIMITATIONS AND THREATS TO VALIDITY

**Collection, Filtering and Analysis of Publications:** In this work, we have focused on in-depth analysis of publications exploring fairness as a software property or conducting fairness analysis via the lens of software engineering (SE). This scope means that we may have missed or filtered out papers that study fairness in other aspects, e.g., as a legal, ethical or transparency concern. Hence, this work is limited to the analysis of fairness property as an SE concern.

Table 7. Details of Open Problems and Future Research Opportunities

| Open Problems | Problem Description | Potential Solutions | Sample Related Work |
|---|---|---|---|
| Fairness Test Metrics and Adequacy | Measuring when fairness testing is sufficient/enough | Design of fairness test metrics and adequacy criteria | [43, 81, 94, 103] [50, 96, 119] |
| Automatic Repair of Biased Classifiers | How to automatically repair biased classifiers to be (less or ) un-biased? | Automatic Program Repair for fairness property | [8, 112, 125] |
| Tooling for Fairness Property Specification | Specifying and engineering fairness properties for learning-based systems | Requirement Engineering tool support for Fairness properties | [14, 117] |
| Unexplored or Poorly Understood Biases | Analyzing rare biases (e.g., age), complex or intersectional biases (e.g., age × gender)? | Fairness Analysis Support for rare, complex or intersectional Biases | [25, 28] |
| Sequential and Long-term Fairness concerns | How to analyse/maintain fairness as the AI system evolves over time? | Techniques to support analysis of sequential and long-term fairness | [146] |
| Human factors in fairness analysis | Evaluating the harm induced by fairness violations to humans/society | Empirical studies of Human Factors in Fairness Analysis | [38, 63, 86] |
| Non-Specific/Holistic mitigation approaches | Designing bias mitigation methods that are agnostic of tasks, domains or datasets | General (i.e., task, domain and dataset -agnostic) bias analysis techniques | [145] |
| Fair Policy, Legalisation, and Compliance | How to design fairness analysis tools for policy makers and compliance officers? | Fairness Analysis Tool Support for Policy and Compliance Analysis | [89, 102] |

**Manual Publication Analysis/Interpretation:** The analysis of the publications studied in this paper are potentially open to human bias since they were manually coded and analyzed. However, to mitigate this threat we ensure that the in-depth analysis and categorization of each paper is validated by at least one other researcher.

# 7 CONCLUSION AND FUTURE OUTLOOK

This paper presents a comprehensive analysis and survey of publications on software fairness. In particular, we study publications that study *fairness as a software property*, works that examine fairness with a software engineering lens, or study how to engineer fair learning-based software systems. We identified several open problems and gaps. Table 7 highlights the details of the open problems identified in this work. Specifically, we highlight open problems in the engineering of fair software systems, including concerns in the areas of fairness testing, verification and empirical evaluation of fairness properties. In the following we discuss the gaps and open problems we elicited from our analysis.

In our study, we found no work *exploring the test adequacy of fairness testing approaches*. Even though there are several papers exploring fairness testing concerns [5, 130, 143] as well as works investigating test metrics for learning-based systems [66, 93, 124, 136], there is still the *problem of measuring test adequacy for fairness testing*. We found *no work investigating test adequacy for fairness concerns*. Invariably, several questions concerning the adequacy of fairness testing remains unanswered and unexplored: When is fairness testing (in)sufficient? How can fairness testing be guided to reduce redundant test cases? Beyond fairness violations, are there other test metrics that indicate (in)sufficient testing has been performed? These are open problems in the area of fairness testing.

Despite the advances in test metrics for traditional and learning-based software, determining the appropriate test metrics or adequacy criteria for fairness evaluation remains an open problem. Researchers have proposed several test adequacy criteria for learning-based software, such as *neuron coverage* [103] and *surprise adequacy* [81]. *Neuron coverage* (from DeepXplore [103]) measures the parts of a learning-based systems that is exercised by a test input, it employs the ratio of neurons whose activation values were above a predefined threshold to measure the diversity of neuron behaviour and guide test generation. Likewise, *surprise adequacy* evaluates the behaviour of learning-based system with respect to their training data by measuring the surprise of an input as the difference in the system's behaviour between the input and the training data, such that a good test input should be sufficiently but not overtly surprising compared to the training data. Other test metrics for learning-based systems include DeepGini [50], Multiple-Boundary Clustering and Prioritization (MCP) [119] and Maximum Probability (MaxP) [96]. Despite the availability of several test metrics and adequacy criteria for functional testing of learning-based systems, there is no *indication or evaluation that demonstrates these test metrics are applicable for the fairness testing of learning-based systems*. The test metric typically employed by the fairness testing literature remains unfair behavior/outputs characterizing fairness violations. We encourage researchers to further investigate this vital challenge of fairness testing.

Although there are several fairness mitigation approaches, *repairing unfair models to fulfill software fairness properties (i.e., to be fair or unbiased) has been hardly explored*, except for few works (e.g., Albarghouthi et al. [8]). Albarghouthi et al. [8] proposed an approach that applies *distribution-guided inductive synthesis* to repair *unfair* ML classifiers with the goal of making them fair, it also verifies that the repaired classifiers is semantically close to the original (unfair) classifier. Their approach formulates the problem as a probability distribution problem, such that the repaired classifier needs to satisfy a probabilistic Boolean expression. Most importantly, this work is one of the few approaches we have found that aims to directly repair ML classifiers to fulfill fairness properties, without model re-training. Similarly, there is *low support for verification of fairness properties* (*see* RQ6). We had observed that verifying, certifying and providing guarantees for software fairness in learning-based software is under-explored.

Additional open problems include the *low number of empirical evaluation of fairness properties, especially as they relate to human factors and societal policies*. There are limited empirical studies studying or measuring the harm caused by unfair software behavior (to humans) and the impact of fairness mitigation on marginalized individuals and communities. Blodgett et al. [23] emphasizes the importance of evaluating the harms induced by unfair (NLP) systems on humans and societies.

Furthermore, there is *a need to develop approaches and tool support that are task, and dataset agnostic*. For instance, we observed that there is a lack of general, non-specific fairness mitigation approaches, most works target a specific domain/task (e.g., NLP, CV) or a specific type of dataset (e.g., structured datasets), with little work that is generally or demonstrably applicable across domains, tasks or datasets (except for few works like ADF [145]).

Besides, *there is a need to provide tools to support the automatic specification and requirements engineering of fairness properties*. Closely related works in this area includes MLCheck [117] and VeriFair [14] which allows to specify and check fairness specifications. However, these tools do not provide support for non-technical experts, e.g., compliance officers and policy makers. Indeed, such tools should support the definition and formalization of specific fairness properties as a socio-technical issue, not just a technical issue. For instance, these tools should allow not only allow to test fairness, but enable compliance officers and policy makers to audit and derive bias-preserving policies (e.g., equity-based solutions like affirmative action [102]), respectively.

In RQ4, we demonstrate that there are some *poorly explored and understood biases, especially in terms of protected attributes*. Generally, we observed that certain tasks, datasets and biases are poorly studied. For instance, protected attributes relating to marital status, sexuality, non-binary gender, and education are poorly explored. Relating to this issue is the fact that the interactions of protected attributes is under-explored, especially in terms of the compounding or intersectional effect of multiple protected attributes, e.g., biases triggered by a combination of attributes (e.g., age × gender). While works like FairVis [28] allow to visualize and analyze intersectional bias, there is little support for specifying, testing and mitigating such biases.

Sequential and long-term fairness concerns remain an open problem [146]. For instance, how do we mitigate fairness properties as the software system evolves? Do the proposed mitigation approaches for one-time fairness analysis scale to sequential or long-term concerns. In addition, there is the socio-technical and ethical concern of fairness analysis: How do we ensure that our fairness mitigation approaches do not induce new and unintended biases? How do our proposed mitigation and analysis approaches translate to real-world intervention for fairness, e.g., in terms of equity-based mitigation approaches typically employed in public policy (e.g., affirmative action [102])?

In summary, we posit that there is a need for socio-technical, human-in-the-loop bias analysis approaches that translate to the mitigation of real-world harms to humans and society. There is a need to provide bias analysis methods to support developers, policy makers and compliance officers. For easy scrutiny, reuse and replication, we provide details of collected papers, our in-depth analysis for each research question and findings:

https://github.com/ezekiel-soremekun/Software-Fairness-Analysis